\documentclass[preprint,12pt]{elsarticle}

\usepackage{amssymb}
\usepackage{amsmath}
\usepackage{array}
\usepackage{epstopdf}

\journal{Journal of Magnetism and Magnetic Materials}

\begin{document}

\begin{frontmatter}



\title{Polaritons dispersion in a composite ferrite-semiconductor structure near gyrotropic-nihility state}

\author[rian]{Vladimir~R.~Tuz\corref{correspond}}
\ead{tvr@rian.kharkov.ua}

\address[rian]{Institute of Radio Astronomy of National Academy of Sciences of Ukraine, \\Kharkiv, Ukraine}
\cortext[correspond]{Institute of Radio Astronomy of National
Academy of Sciences of Ukraine, 4, Mystetstv St., Kharkiv
61002, Ukraine}

\begin{abstract}
In the context of polaritons in a ferrite-semiconductor structure which is influenced by an external static magnetic field, the gyrotropic-nihility can be identified from the dispersion equation related to bulk polaritons as a particular extreme state, at which the longitudinal component of the corresponding constitutive tensor and bulk constant simultaneously acquire zero. Near the frequency of the gyrotropic-nihility state, the conditions of branches merging of bulk polaritons, as well as an anomalous dispersion of bulk and surface polaritons are found and discussed.
\end{abstract}


\end{frontmatter}



\section{Introduction}
\label{sec:intro}

The polaritons are modes of the electromagnetic field coupled to normal modes (eigenwaves) which are inherent to a material and able to interact in a linear manner with the electromagnetic field by virtue of their electrical or magnetic character \cite{Mills_RepProgPhys_1974}. According to the quantum description, polaritons are related to some `quasi-particle' excitations consisting a photon coupled to an elementary excitation like plasmon, phonon, exciton, etc., which bring some polarizability into media. Nevertheless, since polaritons are modes of the electromagnetic field, their description is possible to fulfill on the basis of \textit{macroscopic} Maxwell's equations, where polaritons are considered as modes existing in a bulk material (bulk polaritons) as well as on a medium surface (surface polaritons). In the latter case corresponding boundary conditions must be imposed in order to match the components of the field vectors on both sides of the interface. Therefore, electromagnetic features of these bulk and surface modes in matter turn out to be closely related to the material properties of a medium, and, in particular, to the resonant states in the frequency dependence of its macroscopic dielectric function \cite{Pitarke_RepProgPhys_2007}. In such approach, the polaritons are completely determined by the characteristic frequencies of the relative permittivity of the medium.

Furthermore, in the case of magneto-optically active media the problem of polaritons is mostly related to two distinct considerations of gyroelectric media with the (surface) magnetoplasmon and gyromagnetic media with the (surface) magnon which involve the media characterization with either permittivity or permeability tensor having antisymmetric off-diagonal parts \cite{Burstein_PhysRevLett_1972, Burstein_JPhysC_1973, Burstein_PhysRevB_1974, Rapoport_NewJPhys_2005, Tagiyeva_Superlattices_2007, Hu_Nanophot_2015}\footnote{For the first reading on classification of polaritons in magnetic media we refer to a comprehensive review \cite{Kaganov_PhysUsp_1997}.}. This distinction is convenient due to the physical mechanisms that cause corresponding resonant states (e.g. vibration of molecules in dielectrics or magnetization in ferrites). Thus, characteristic frequencies of permittivity are confined to the optical range (sometimes, in ion crystals, to the infrared one), whereas those of permeability are in the microwave range (occasionally, in antiferromagnets, in the submillimeter range).

Although characteristic frequencies of permittivity and permeability normally lie far from each other, it is possible to find exceptions to this rule. In particular, bigyrotropic media can be implemented artificially by properly combining together gyroelectric and gyromagnetic materials in a certain proportion or admixing inclusions of a special kind into a host medium. As relevant examples  layered magnetic-semiconductor heterostructures \cite{Ivanov_JMMM_2006, Wu_JPhysCondMatt_2007, Tarkhanyan_PSSb_2008, Shramkova_PIERM_2009} and the magnetized electron-ion plasma consisting micron ferrite grains \cite{Rapoport_PhysPlasmas_2010} can be mentioned. Such \textit{gyroelectromagnetic} composites are now usually considered within the theory of metamaterials, in the framework of which they are widely discussed from the viewpoint of achieving negative refraction and backward wave propagation\footnote{We consider a backward wave as a wave in which the direction of the Poynting vector is opposite to that of its phase velocity \cite{Lindell_MOP_2001}.}.

Further,  the role of bigyrotropy inherent to gyroelectromagnetic media in the spectral features of polaritons is elucidated in \cite{Tarkhanyan_PhysicaB_2010, Tarkhanyan_JMMM_2010, Bulgakov_EurPhysJApplPhys_2014}, where necessary conditions for negative refraction and anomalous dispersion are found out in the (anti)ferromagnet-semiconductor superlattices (e.g., in MnF$_2$-CdTe nanostructures) being under an action of an external static magnetic field. It is shown that the applied magnetic field essentially changes the spectrum of polaritons in such composite media, increases the number of spectral branches, while the characteristic frequencies become to be strongly dependent on the magnetic field. As a result, the dispersion features of polaritons in different ranges of the external magnetic field are quite diverse, and not only the number but also the limits of the frequency ranges with anomalous dispersion appear to be very sensitive to the strength of the magnetic field.

As a gradual development of this theory, in the present paper we study particular spectral features of polaritons that appear in a gyroelectromagnetic medium in the frequency band near a certain extreme state in the material dispersion, when relative permittivity and relative permeability of the medium simultaneously turn into zero. In fact, such a peculiarity is discussed within the conceptions of `nihility' \cite{Lakhtakia_IntIRMMWaves_2002, Tretyakov_JEMWA_2003, Tuz_PIER_2010} and `epsilon-and-mu-near-zero' media \cite{Alu_PhysRevB_2007, Engheta_NatCommun_2014}\footnote{We believe that these two terms `nihility' and `epsilon-and-mu-near-zero' define the same physical essence and the difference between them is only in their names. Therefore, further we use the term `nihility' by virtue of personal preferences.}, and is defined in \cite{Shen_PhysRevB_2006, Tuz_PIERB_2012, Tuz_JOpt_2015, Tuz_Book_2016} as a `gyrotropic-nihility' state. Thus, in the \textit{Faraday geometry} the gyrotropic-nihility state is found can be achieved in a composite ferrite-semiconductor superlattice in the microwave part of spectrum near the frequencies of ferromagnetic and plasma resonances. In this case real parts of diagonal elements of both complex effective permittivity and complex effective permeability tensors of such artificial medium simultaneously acquire zero, while the off-diagonal elements appear to be non-zero quantities. Here we focus on the search of conditions for existence of the gyrotropic-nihility state in the gyroelectromagnetic system being in the \textit{Voigt geometry} from the viewpoint of the polaritons problem. For the bulk polaritons we revisit the results of \cite{Kaganov_PhysStatSolB_1990}, where the particular extreme state when permittivity and permeability of media simultaneously turn into zero is referred to `resonant polaritons'\footnote{Here again a question about the terminology arises. The term `resonant polaritons' is also introduced in \cite{Tassone_IlNuCimD_1990} for the resonant state of the Breit-Wigner type in the scattering amplitude of the electromagnetic waves in quantum wells.}. Besides, spectral properties of the surface polaritons are also studied.

\section{Dispersion relations for bulk and surface polaritons}
\label{sec:dispersion}

\begin{figure}[htbp]
\centerline{\includegraphics[width=0.5\columnwidth]{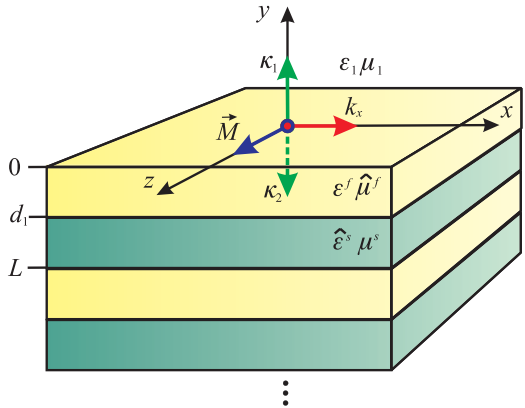}}
\caption{A composite finely-stratified ferrite-semiconductor structure, which is influenced by an external static magnetic field in the Voigt geometry.} \label{fig:fig1}
\end{figure}

Thereby, further in this paper we study polaritons dispersion features in a \textit{semi-infinite} stack of identical double-layer slabs (unit cells) periodically arranged along the $y$-axis (figure~\ref{fig:fig1}). Each unit cell is constructed by juxtaposition together of ferrite (with constitutive parameters $\varepsilon^f$, $\hat \mu^f$) and semiconductor (with constitutive parameters $\hat \varepsilon^s$, $\mu^s$) layers with thicknesses $d_f$ and $d_s$, respectively. The stack possesses a periodic structure (with period $L = d_f + d_s$) that fills half-space $y<0$ and adjoins an isotropic medium (with constitutive parameters $\varepsilon_1$, $\mu_1$) occupying half-space $y>0$. We consider that the structure is a \textit{finely-stratified} one, i.e. its characteristic dimensions $d_f$, $d_s$ and $L$ are all much smaller than the wavelength in the corresponding layer $d_f\ll \lambda$, $d_s \ll \lambda$, and period $L \ll \lambda$  (the long-wavelength limit). An external static magnetic field $\vec M$ is directed along the $z$-axis. It is supposed that the strength of this field is high enough to form a homogeneous saturated state of ferrite as well as semiconductor subsystems.

Since all characteristic dimensions $d_f$, $d_s$ and $L$ of the structure under investigation satisfy the long-wavelength limit, the standard homogenization procedure from the effective medium theory (see, \cite{Agranovich_SolidStateCommun_1991, Eliseeva_TechPhys_2008}, and \ref{sec:effmedium})\footnote{As an alternative to \cite{Agranovich_SolidStateCommun_1991, Eliseeva_TechPhys_2008} a powerful approach based on $4\times 4$-transfer matrix formalism \cite{Lakhtakia_Naturforsch_1991, Lakhtakia_Electromagn_2009, Lakhtakia_book_2013} should be mentioned. It enables to perform the homogenization procedure and obtain a solution of the boundary-value problem with respect to polaritons in the most general case of linear bianisotropic materials. Nevertheless, in this paper we consciously made a choice in favor of the method \cite{Agranovich_SolidStateCommun_1991}, since it makes possible to obtain an explicit expressions for all components of effective permeability and effective permittivity tensors, which are further needed for solving an optimization problem and identification of asymptotes of polariton branches.} is applied in order to derive averaged expressions for effective constitutive parameters of the structure. In this way, the finely-stratified multilayered system is approximately represented as a \textit{biaxial} anisotropic (gyroelectromagnetic) uniform medium, whose  first optical axis is directed along the structure periodicity, while the second one coincides with the direction of the external static magnetic field $\vec M$. Therefore, the resulting composite medium is characterized with the tensors of relative effective permittivity $\hat\varepsilon_{eff}$ and relative effective permeability $\hat\mu_{eff}$, whose expressions derived via underlying constitutive parameters of ferrite ($\varepsilon^f$, $\hat \mu^f$) and semiconductor ($\hat \varepsilon^s$, $\mu^s$) subsystems one can find in \ref{sec:constpar}.

Let us further consider a surface wave which propagates along the interface of an isotropic dielectric and a composite structure (we distinguish these media with the index $j=1,2$). According to figure~\ref{fig:fig1} the $y$-axis is perpendicular to the interface between the media, and the wave propagates along the $x$-axis with a wavenumber $k_x$. The attenuation of the wave in both positive and negative directions along the $y$-axis is defined by quantities $\kappa_j$. The static magnetic field $\vec M$ is directed along the $z$-axis, so the system is considered to be in the Voigt geometry.

In a \textit{general} form, the electric and magnetic field vectors $\vec E$ and $\vec H$ used here are represented as \cite{Lakhtakia_book_2013} 
\begin{equation}
\vec P^{(j)}= \vec p^{(j)}\exp\left(i k_x x\right)\exp\left(\mp
\kappa_j y\right), \label{eq:inc}
\end{equation}
where a time factor $\exp\left(-i \omega t\right)$ is also supposed and omitted, and sign `$-$' is related to the fields in the upper medium ($y>0$, $j=1$) and sign `$+$' is related to the fields in the composite medium ($y<0$, $j=2$), respectively, which provide required wave attenuation along the $y$-axis.

Involving a pair of the divergent Maxwell's equations $\nabla \cdot \vec D=0$ and $\nabla \cdot \vec B=0$ in the form 
\begin{equation}
\nabla \cdot \vec Q^{(j)}=\nabla \cdot \left(\hat g^{(j)} \vec
P^{(j)}\right)=0, \label{eq:divergent}
\end{equation}
where $\hat g^{(1)}$ is related to the upper medium as a tensor quantity with $\varepsilon_1$ or $\mu_1$ on its main diagonal and zeros elsewhere (i.e., $\hat g^{(1)}=g_1 \hat I$, where $\hat I$ is the identity tensor), and $\hat g^{(2)} = \hat g_{eff}$ is related to the composite medium, one can immediately arrive to the relations between the field components in the upper ($y>0$) and composite ($y<0$) medium as follows
\begin{equation}
P^{(1)}_y = i\frac{k_x}{\kappa_1}P^{(1)}_x,~~~P^{(2)}_y = i\frac{\kappa_2 g_{xy}-i k_x g_{xx}}{i \kappa_2 g_{xx} - k_x g_{xy}}P^{(2)}_x,
\label{eq:fieldcomp}
\end{equation}
from which the field polarization at an arbitrary fixed point of both media can be determined.

Taking into consideration continuity of components $P^{(j)}_x$ and $Q^{(j)}_y$ at the interface $y=0$ between the media
\begin{equation}
g_1\frac{P^{(1)}_y}{P^{(1)}_x} = g_{yy}\frac{P^{(2)}_y}{P^{(2)}_x}-g_{xy},
\label{eq:continuity}
\end{equation}
and relations (\ref{eq:fieldcomp}), the dispersion equation for the \textit{surface} polaritons at the interface between isotropic dielectric and bigyrotropic (gyroelectromagnetic) semi-infinite media is obtained as
\begin{equation}
\kappa_2 g_1 + \kappa_1\left(g_{xx}+\frac{g_{xy}^2}{g_{yy}}\right)+i k_x g_1\frac{g_{xy}}{g_{yy}}=0. 
\label{eq:dispsurface}
\end{equation}

From a pair of the curl Maxwell's equations $\nabla\times\vec E = i
k_0 \vec B$ and $\nabla\times\vec H = -i k_0 \vec D$, where
$k_0=\omega/c$ is the free space wavenumber, in a standard way the
wave equation can be obtained in the form:
\begin{equation}
\nabla \times \nabla \times\vec P^{(j)} + k_0^2 \hat \varsigma^{(j)} \vec P^{(j)} = 0,
\label{eq:waveeq}
\end{equation}
where $\hat \varsigma^{(j)}$ is introduced as the product of $\hat \mu^{(j)}$ and $\hat \varepsilon^{(j)}$ made in the appropriate order.

Substituting field components (\ref{eq:inc}) into the wave equation
(\ref{eq:waveeq}) allows us to derive a relation between the field
components in both the isotropic dielectric and composite media:
\begin{equation}
\left( {\begin{matrix}
k_0^2\varsigma^{(j)}_{xx} + \kappa_j^2       & k_0^2\varsigma^{(j)}_{xy} - i k_x \kappa_j^2 \cr 
k_0^2\varsigma^{(j)}_{yx} - i k_x \kappa_j^2 & k_0^2\varsigma^{(j)}_{yy} - k_x^2 \cr
\end{matrix}
} \right) \left( {\begin{matrix} P^{(j)}_x \cr P^{(j)}_y \cr
\end{matrix}
} \right) =0,\label{eq:fromwaveeq1}
\end{equation}
\begin{equation}
\left(k_0^2\varsigma^{(j)}_{zz} - k_x^2+\kappa_j^2\right)P^{(j)}_z = 0.
\label{eq:fromwaveeq2}
\end{equation}
From (\ref{eq:fromwaveeq1}) and (\ref{eq:fromwaveeq2}) it follows
that in the considered system the TE mode $\{H_x,~H_y,~E_z\}$ is not
coupled to the TM mode $\{E_x,~E_y,~H_z\}$.

Direct substitution of corresponding values into (\ref{eq:fromwaveeq1}) for upper ($\varsigma^{(1)}_{xx}=\varsigma^{(1)}_{yy}=\varsigma^{(1)}_{zz}=\varsigma_1$, $\varsigma^{(1)}_{xy}=\varsigma^{(1)}_{yx}=0$) and composite ($\varsigma^{(2)}_{mn}=\varsigma_{mn}$) media, and expanding the determinant, expressions related to the attenuation functions $\kappa_j$ are obtained in the form:
\begin{equation}
k_x^2-\kappa_1^2-k_0^2\varsigma_1=0. \label{eq:kappa1}
\end{equation}
\begin{equation}
k_x^2\varsigma_{xx} - \kappa_2^2\varsigma_{yy} - k_0^2\varepsilon_{yy}\mu_{yy}\varepsilon_{v}\mu_{v} - i k_x\kappa_2(\varsigma_{xy}+\varsigma_{yx}) =0,
\label{eq:kappa2}
\end{equation}
where $\mu_{v} = \mu_{xx}+\mu_{xy}^2/\mu_{yy}$ and $\varepsilon_{v} = \varepsilon_{xx}+\varepsilon_{xy}^2/\varepsilon_{yy}$ are the Voigt relative permeability and relative permittivity, which can be considered as the bulk magnetic and dielectric constants.

From the curl Maxwell's equations involving (\ref{eq:inc}) and above introduced the Voigt relative permeability $\mu_v$ and relative permittivity $\varepsilon_v$ the dispersion law for the \textit{bulk} polaritons in the composite medium follows as solutions of two separate equations related to the TE mode and the TM mode, respectively,
\begin{equation}
k_x^2\mu_{xx} - \mu_{yy}(k_0^2\varepsilon_{zz}\mu_v+\kappa_2^2)=
0,~~~k_x^2\varepsilon_{xx} -
\varepsilon_{yy}(k_0^2\mu_{zz}\varepsilon_v+\kappa_2^2) = 0,
\label{eq:dispbulk}
\end{equation}
handled in the form of functions of wavenumber $k_x$ versus $k_0$, since  $\mu_{mn}$, $\varepsilon_{mn}$,  $\mu_v$ and $\varepsilon_v$ are all functions of frequency $\omega$ and thus they are functions of $k_0$.

Combining together relations (\ref{eq:kappa1}), (\ref{eq:kappa2}) and dispersion equation (\ref{eq:dispsurface}) with appropriate substitution of $\mu \to g$ for the TE mode, and $\varepsilon \to g$ for the TM mode gives us the closed system of equations whose solutions for the wavenumber $k_x$ versus $k_0$ completely determine the spectral characteristics of the surface polaritons. Consistently getting rid of the attenuation functions $\kappa_j$ from the dispersion equation (\ref{eq:dispsurface}), it can be expanded into the biquadratic equation with respect to $k_x^2$:
\begin{equation}
Ak_x^4 + Bk_x^2+C=0, \label{eq:biquadratic}
\end{equation}
where $A= Y^2+W^2$, $\quad B=-k_0^2(2VY+\varsigma_1W^2)$, $\quad C=k_0^4V^2$, $\quad V=\varepsilon_{yy}\mu_{yy}\varepsilon_v\mu_v - \varsigma_1\varsigma_{yy}\tilde g_v^2$, $\quad Y=\varsigma_{xx}+(\varsigma_{xy}+\varsigma_{yx})\tilde g_{xy}+(\tilde g_{xy}^2 -\tilde g_v^2)\varsigma_{yy}$, $\quad W=(\varsigma_{xy}+\varsigma_{yx}+2\varsigma_{yy}\tilde g_{xy})\tilde g_v$, and $\tilde g_v = g_v/g_1$, $\quad \tilde g_{xy} = g_{xy}/g_{yy}$, whose solution is trivial:
\begin{equation}
k_x^2=\frac{-B\pm\sqrt{B^2-4AC}}{2A}. \label{eq:solutionbi}
\end{equation}
From four roots of (\ref{eq:solutionbi}) those must be selected which satisfy the physical conditions, namely, wave attenuation as it propagates, that imposes restrictions on the values of $\kappa_j$ whose real parts must be positive quantities.

\section{Analysis of polaritons dispersion conditions}
\label{sec:conditions}

From the form of dispersion equations for bulk (\ref{eq:dispbulk}) and surface (\ref{eq:biquadratic}) polaritons it is obvious that their spectral features substantially depend on the dispersion characteristics of components of both effective permeability tensor and effective permittivity tensor of the resulting finely-stratified structure. Moreover, since the components of these tensors are all functions of the frequency, external magnetic field  strength, layers' thicknesses, and physical properties of the materials forming the superlattice, the regions of polaritons existence are determined by the choice of the values of the corresponding quantities. So we are dealing here with a multiparameter problem which requires to be optimally solved in order to achieve the desired dispersion features of polaritons.

Imposing $\kappa_2=0$ into (\ref{eq:dispbulk}) we obtain the dispersion equations 
\begin{equation}
k_x^2 =
k_0^2\varepsilon_{zz}\mu_v\mu_{yy}\mu_{xx}^{-1},~~~k_x^2=k_0^2
\mu_{zz}\varepsilon_v\varepsilon_{yy}\varepsilon_{xx}^{-1},
\label{eq:dispbulk2}
\end{equation}
whose solutions outline the regions of existence of bulk polaritons. As it is defined in \cite{Lakhtakia_IntIRMMWaves_2002}, a `nihility' state implies simultaneous equality to zero of relative permittivity and relative permeability of a medium. Therefore, in the considered composite structure, in the context of the bulk polaritons (\ref{eq:dispbulk2}), such a state corresponds to a particular frequency, where  $\varepsilon_{zz}$, $\mu_v$ and $\mu_{zz}$, $\varepsilon_v$ simultaneously acquire zero for the TE mode and the TM mode, respectively. Since this peculiarity is obviously associated with the media gyrotropy we distinguish it as the `gyrotropic-nihility' state\footnote{We should emphasize that in \cite{Kaganov_PhysStatSolB_1990} the term `resonant polaritons' is introduced for a \textit{hypothetical} isotropic medium in which characteristic frequencies (either resonance or antiresonance) of permittivity and permeability coincide. In this paper we demonstrate that the discussed extreme state when both relative permittivity and relative permeability simultaneously turn into zero can be achieved in a \textit{real} structure with a particular type of anisotropy (gyrotropy). So we use the term `gyrotropic-nihility' in order to distinguish exactly this state among other ones.}. From the dispersion characteristics of the constitutive parameters one can conclude that in the structure under study the gyrotropic-nihility state for the TM mode is not achievable at all, since the longitudinal component $\mu_{zz}$ of the effective permeability tensor is a constant nonzero quantity in the whole frequency band of interest. From the physical point of view, it is a consequence of the fact that the magnetic field vector in the TM mode is parallel to the external magnetic field, which results in the absence of its interaction with the magnetic subsystem.

\begin{figure}[htbp]
\centerline{\includegraphics[width=1.0\columnwidth]{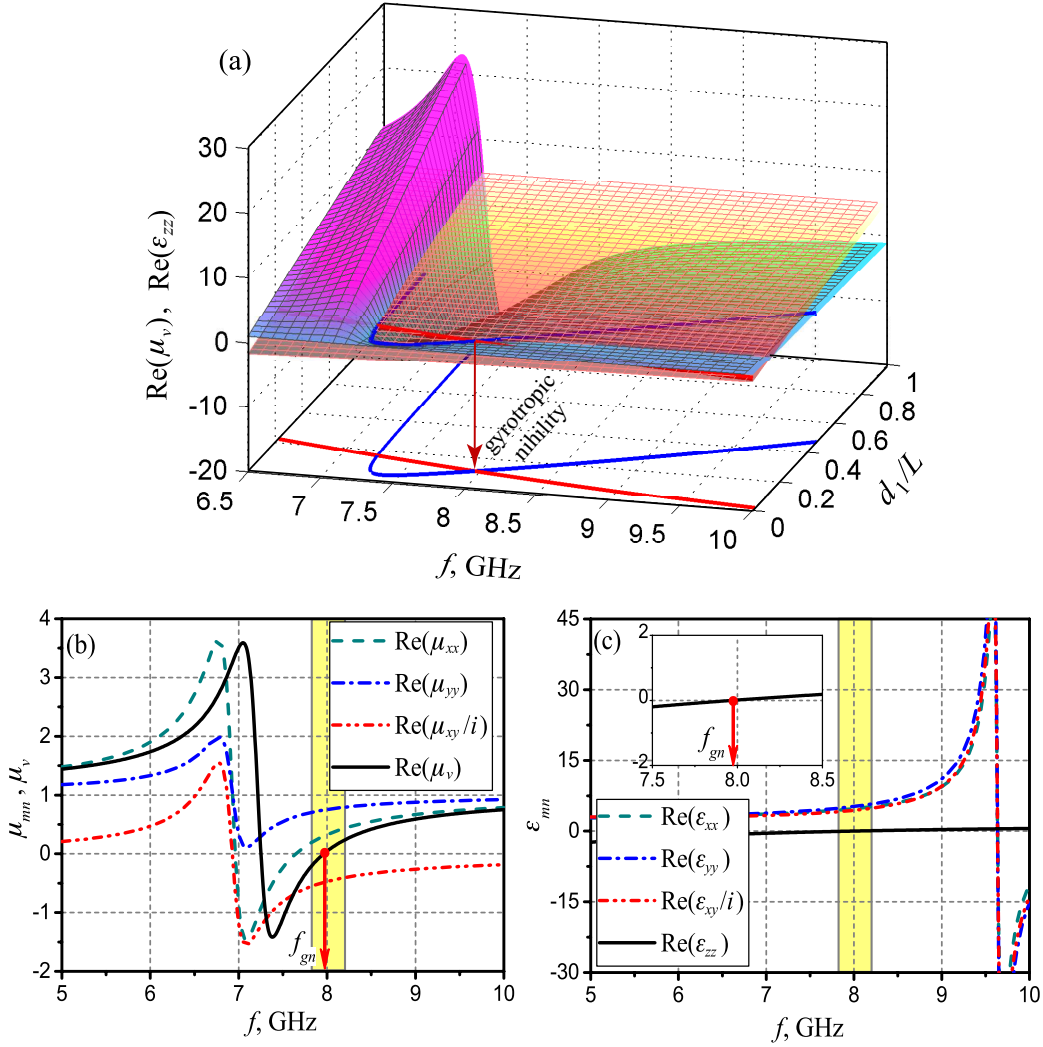}}
\caption{(a) Two surfaces depict behaviors of real parts of relative
permeability (purple surface) and  relative permittivity (orange
surface) versus frequency and the ratio of the layers' thicknesses.
The blue and red curves plotted on the surfaces show the conditions
Re$(\mu_v)=0$ and Re$(\varepsilon_{zz})=0$, respectively. Curves at
the bottom are just projections and are given for an illustrative
purpose. Dispersion curves of the tensor components of (b) effective
permeability and (c) effective permittivity at particular structure
parameters ($\delta_1=0.117$, $\delta_2=0.883$, $L=1$~mm) for which
the gyrotropic-nihility state exists. For the ferrite layers, under saturation magnetization of 2000~G,
parameters are: $\omega_0/2\pi=4.2$~GHz, $\omega_m/2\pi=8.2$~GHz,
$b=0.02$, $\varepsilon^f=5.5$. For the semiconductor layers, parameters are: $\omega_p/2\pi=10.5$~GHz, $\omega_c/2\pi=9.5$~GHz,
$\nu/2\pi=0.05$~GHz, $\varepsilon_l=1.0$, $\mu^s=1.0$.} \label{fig:fig2}
\end{figure}

Thus, the gyrotropic-nihility state can be achieved only for the TE
mode, and the search of conditions for its existence implies solving
an optimization problem, where $\varepsilon_{zz}$ and $\mu_v$ are
subjected to zero in the objective function. During the solution of
this problem both the magnetic field strength (and, accordingly,
characteristic frequencies of the constitutive parameters) and
structure period are fixed, and the search for the frequency where
the gyrotropic-nihility state appears is proceeded via altering the
layers' thicknesses within the period. The graphical solution of the
discussed optimization problem is depicted in figure \ref{fig:fig2},
where the resolved gyrotropic-nihility state is distinguished by
arrows. In the structure under consideration this state is found to
be at a particular frequency $f_{gn}=7.97$~GHz which corresponds to
the following geometric factors: $\delta_1=d_f/L=0.117$; $\delta_2=d_s/L=0.883$;
$L/\lambda \approx 3\times10^{-2}$.

\begin{figure}[htbp]
\centerline{\includegraphics[width=0.6\columnwidth]{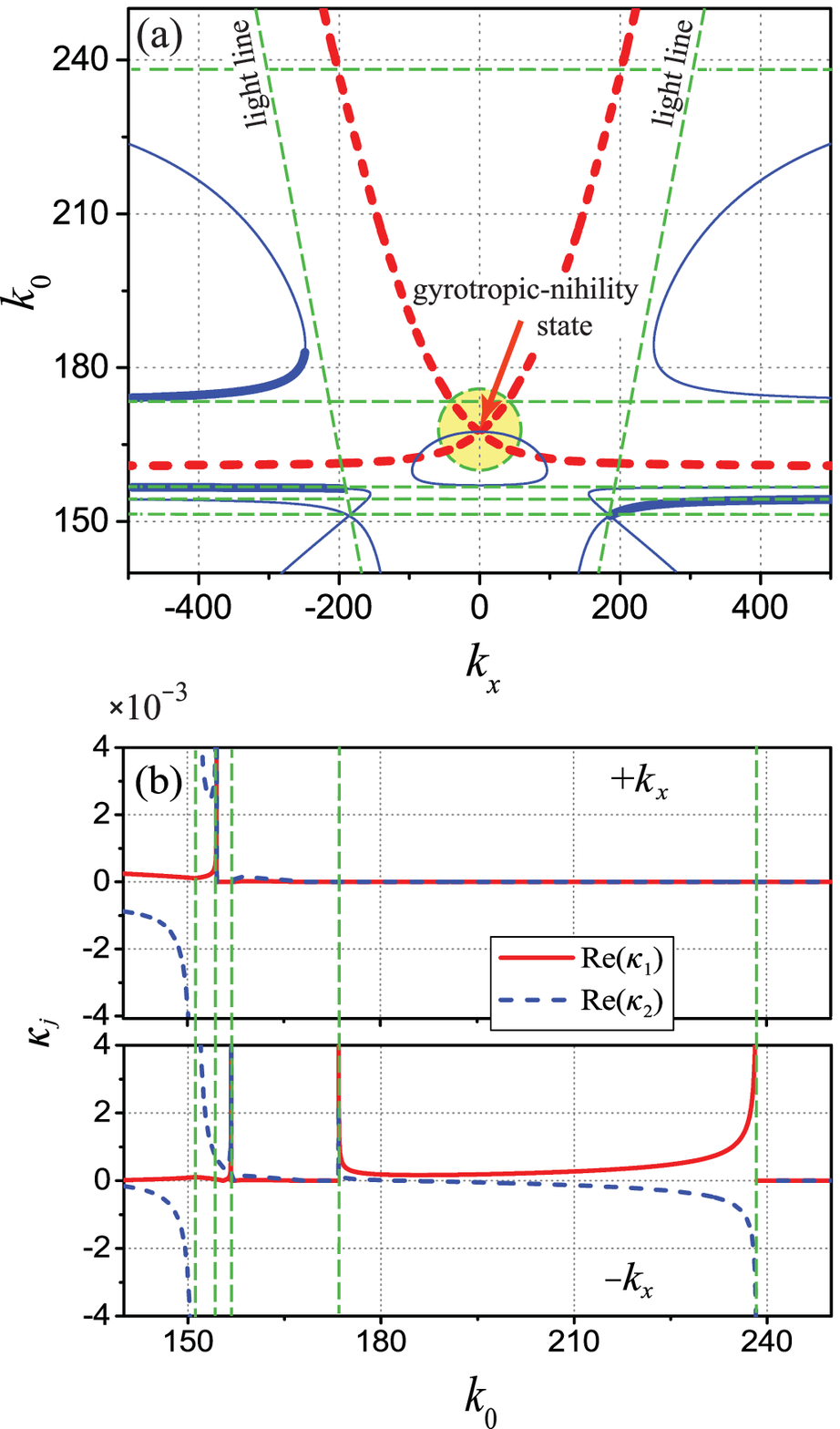}}
\caption{(a) Dispersion characteristics of the bulk (red dash lines) and surface (blue thick solid lines) polaritons. Blue thin solid lines depict four solutions of the dispersion equation for the surface polaritons. Green dash lines depict the asymptotic conditions. (b) Frequency dependencies of the attenuation functions $\kappa_j$ in the positive (upper panel) and negative (bottom panel) ranges of $k_x$. $\delta_1=0.117$, $\delta_2=0.883$, $L=1$~mm. Other parameters of the ferrite and semiconductor layers are the same as in figure~\ref{fig:fig2}.} \label{fig:fig3}
\end{figure}

Since further our goal is to elucidate the dispersion laws of the bulk polaritons (which are in fact eigenwaves), we are interested in \textit{real} solutions of the first equation in (\ref{eq:dispbulk2}) for the TE mode. In order to find the real solutions related to eigenwaves, absence of losses in constitutive parameters of the underlying layers is supposed, i.e. we put $b=0$ and $\nu=0$ in tensors (\ref{eq:gfs}) for ferrite and semiconductor subsystems, respectively. These solutions are depicted in figure \ref{fig:fig3}(a) with the red dashed lines.

In the systems which stand out with two characteristic frequencies, where one resonance exists for permittivity and other one exists for permeability, the dispersion law of the bulk polaritons has three branches \cite{Tarkhanyan_JMMM_2010}, and typically these branches are separated from one another by some forbidden bands. In our case, the resonance of the relative permittivity $\varepsilon_{zz}$ is at zero frequency, therefore, there are only two branches in the dispersion law without any forbidden band between them. Remarkably, the branches merge exactly at the frequency of the gyrotropic-nihility state, which is marked out in figure \ref{fig:fig3}(a) with a circle. The upper branch is restricted by the light lines, wherein the bottom branch manifests an anomalous dispersion and is restricted by the line at which $\mu_{xx} = 0$. It should be noted that although the dispersion curves of the bulk polaritons are calculated for an idealized lossless structure, in a real system the losses are expected to be small near the frequency of the gyrotropic-nihility state, since zero states of $\varepsilon_{zz}$ and $\mu_v$ are both far from the corresponding frequencies of the plasma and ferromagnetic resonances.

Still assuming that there are no losses in the constitutive layers, four solutions of dispersion equation (\ref{eq:biquadratic}) with respect to the propagation constant $k_x$ of the surface polaritons are calculated and presented in figure~\ref{fig:fig3}(a) with blue thin solid lines. Importantly, since the dispersion equation consists of a term which is linearly depended on $k_x$ (see the last term in (\ref{eq:kappa2})), the spectral characteristics of the surface polaritons in the structure under study possess the nonreciprocal nature, i.e. $k_0(k_x)\ne k_0(-k_x)$. Therefore, the appropriate root branches between these four solutions should be properly selected in order to ensure that they are physically correct.

As was already mentioned, among these four solutions those must be chosen which provide the wave attenuation as it propagates. It imposes the restriction that both attenuation functions $\kappa_j$ ($j=1,2$) simultaneously must be real positive quantities. The corresponding curves of $\kappa_j$ are presented in figure~\ref{fig:fig3}(b), and taking into account their values for either positive or negative range of $k_x$, the physically correct root branches are  highlighted in figure~\ref{fig:fig3}(a) with blue thick solid lines. One can see that there is one root branch in the range of positive $k_x$, while there are two root branches in the range of negative $k_x$. From equation (\ref{eq:solutionbi}) it is evident that the asymptotic limits of all these root branches are defined by the condition $A = 0$ (i.e. $Y^2+W^2=0$, implying for the lossless system $Y$ is a real number, while $W$ is an imaginary number, therefore, $A$ is always a real number, too). As a result, the surface polariton
branches are restricted by two asymptotic conditions: $Y-iW=0$ and $Y+iW=0$ in positive and negative ranges of $k_x$, respectively.

Furthermore, in traditional systems the regions of existence of bulk and surface polaritons do not overlap, and the surface polaritons can propagate in the gaps between the regions where the bulk polaritons exist \cite{Mills_RepProgPhys_1974}. In the considered case, two bottom root branches confirm this rule. Nevertheless, the upper root branch in the range of negative $k_x$ manifests a violation of this rule, demonstrating situation when the bulk and surface polaritons exist in the same frequency range, but with different wavevectors, and this particular root branch of the surface polariton possesses an anomalous dispersion.

\section{Conclusions}

To conclude, in this paper the dispersion relations for both bulk and surface polaritons in a finely-stratified ferrite-semiconductor structure which is influenced by an external static magnetic field in the Voigt geometry are derived. From the dispersion equation related to the bulk polaritons the conditions for  existence of a particular extreme gyrotropic-nihility state are identified through solving a multiparameter optimization problem. 

From analysis of the dispersion curves related to the bulk polaritons it is found that in their spectral characteristics there are two branches without any forbidden bands between them, and one of these branches manifests an anomalous dispersion.  Remarkably, these two branches merge exactly at the frequency of the gyrotropic-nihility state. The spectral features of the surface polaritons are also elucidated. These spectra possess the nonreciprocal nature with different number of root branches in the positive and negative ranges of $k_x$. Furthermore, one root branch in the negative range of $k_x$ manifests an anomalous dispersion.

It is expected that the real system should demonstrate a low level of losses for the bulk and surface polaritons near the frequency of the gyrotropic-nihility state, since zero states of both relative permittivity and relative permeability lie far from the corresponding frequencies of the plasma and ferromagnetic resonances. Verification of this statement requires additional elaboration.

Nevertheless, from a theoretical viewpoint, the existence of the discussed effects in the polariton spectra is not in doubt, but the experimental verification is still a challenging task. It requires accurate choosing constitutive materials, clear understanding their characteristic frequencies, solving an optimization problem for search magnetic filed strength, corresponding thicknesses and number of layers within the practical system, but nevertheless, we believe in its feasibility. Our confidence in success  is confirmed by the facts that in semiconductors the density of the free charge carriers strongly depends on the temperature, as well as on the type of impurities and the doping level and thus can be varied in a wide interval, therefore, the necessary values of the effective plasma frequency can be achieved and adjusted to the properties of the magnetic subsystem.

\section{Acknowledgement}

This work was supported by National Academy of Sciences of Ukraine with Program `Fundamental issues of creation of new nanomaterials and nanotechnologies' for 2015-2019 years, Project no.~13/15-H.

\appendix

\section{Effective medium theory}
\label{sec:effmedium}
Taking into account the smallness of the layers' as well as period's thicknesses compared to the wavelength ($d_f\ll \lambda$, $d_s \ll \lambda$, and $L\ll\lambda$), the effective medium theory is engaged in order to derive expressions for the averaged constitutive parameters of the periodic structure under investigation. Following the method used earlier for deriving the components of the effective permittivity tensor \cite{Agranovich_SolidStateCommun_1991, Bulgakov_PhysSolidState_2012} and the effective permeability tensor \cite{Eliseeva_TechPhys_2008, Eliseeva_JMMM_2010} for multilayered systems (superlattices) consisted of anisotropic constituents we can formulate the expressions of $\hat\varepsilon_{eff}$ and $\hat\mu_{eff}$ in a \textit{general} form.

In this way, constitutive equations $\vec B = \hat\mu \vec H$ and $\vec D = \hat\varepsilon \vec E$ for ferrite $(0<z<d_f)$ and semiconductor $(d_f<z<L)$ layers, respectively, can be written as follow:
\begin{equation}
Q^{(j)}_m =\sum_n g^{(j)}_{mn}P^{(j)}_n, \label{eq:constit}
\end{equation}
where $\vec  Q$ is substituted for the magnetic and electric flux densities $\vec B$ and $\vec D$; $\vec P$ is substituted for the magnetic and electric field strengths $\vec H$ and $\vec E$; $g$ is substituted for permeability and permittivity $\mu$ and $\varepsilon$; the superscript ($j$) is introduced to distinguish between ferrite $(f \to j)$ and semiconductor $(s \to j)$ layers; subscripts $m$ and $n$ are substituted to iterate over corresponding indexes of the tensor quantities in Cartesian coordinates.

In the chosen problem geometry, the $y$ axis is perpendicular to the interfaces between the layers within the structure, and, therefore, components $P^{(j)}_x$, $P^{(j)}_z$, and $Q^{(j)}_y$ are continuous. Thus, the particular component $P^{(j)}_y$  can be expressed  from equation (\ref{eq:constit}) in terms of the continuous components of the field
\begin{equation}
P^{(j)}_y= - \frac{g^{(j)}_{yx}}{g^{(j)}_{yy}}P^{(j)}_x +
\frac{1}{g^{(j)}_{yy}}Q^{(j)}_y -
\frac{g^{(j)}_{yz}}{g^{(j)}_{yy}}P^{(j)}_z, \label{eq:induction_1}
\end{equation}
and substituted into equations for components $Q^{(j)}_x$ and $Q^{(j)}_z$:
\begin{equation}
\begin{split}
&Q^{(j)}_x=\left(g^{(j)}_{xx} - \frac{g^{(j)}_{xy}g^{(j)}_{yx}}
{g^{(j)}_{yy}}\right) P^{(j)}_x +
\frac{g^{(j)}_{xy}}{g^{(j)}_{yy}}Q^{(j)}_y + \left(g^{(j)}_{xz} -
\frac{g^{(j)}_{xy}g^{(j)}_{yz}}{g^{(j)}_{yy}}\right)
P^{(j)}_z, \\
&Q^{(j)}_z=\left(g^{(j)}_{zx} -
\frac{g^{(j)}_{zy}g^{(j)}_{yx}}{g^{(j)}_{yy}}\right) P^{(j)}_x +
\frac{g^{(j)}_{zy}}{g^{(j)}_{yy}}Q^{(j)}_y + \left(g^{(j)}_{zz} -
\frac{g^{(j)}_{zy}g^{(j)}_{yz}}{g^{(j)}_{yy}}\right) P^{(j)}_z.
\end{split}
\label{eq:induction_2}
\end{equation}
Then these obtained relations (\ref{eq:induction_1}) and (\ref{eq:induction_2}) are used for the fields averaging \cite{Agranovich_SolidStateCommun_1991}.

Since in the long-wavelength limit the fields $\vec P^{(j)}$ and $\vec Q^{(j)}$ inside the layers are considered to be constant, the \textit{averaged} (Maxwell) fields $\langle\vec Q \rangle$ and $\langle\vec P\rangle$ can be determined by the equalities 
\begin{equation}
\langle \vec P\rangle=\frac{1}{L}\sum_j \vec P^{(j)} d_j, ~~~~~
\langle \vec Q\rangle=\frac{1}{L}\sum_j \vec Q^{(j)} d_j.
\label{eq:mean_1}
\end{equation}
In view of the above discussed continuity of components $P^{(j)}_x$, $P^{(j)}_z$, and $Q^{(j)}_y$, it follows that
\begin{equation}
\langle P_x \rangle = P^{(j)}_x, ~~~\langle P_z \rangle = P^{(j)}_z,~~~ \langle Q_y\rangle= Q^{(j)}_y, \label{eq:conditions}
\end{equation}
and on the basis of equations (\ref{eq:induction_1}) and (\ref{eq:induction_2}), the relations between the averaged fields components are obtained as:
\begin{equation}
\begin{split}
\langle Q_x\rangle&=\alpha_{xx}\langle P_x\rangle +
\gamma_{xy}\langle Q_y\rangle + \alpha_{xz}\langle P_z\rangle, \\
\langle P_y\rangle&=\beta_{yx}\langle P_x\rangle +
\beta_{yy}\langle Q_y\rangle + \beta_{yz}\langle P_z\rangle, \\
\langle Q_z\rangle&=\alpha_{zx}\langle P_x\rangle +
\gamma_{zy}\langle Q_y\rangle + \alpha_{zz}\langle P_z\rangle,
\end{split}
\label{eq:mean_2}
\end{equation}
where $\alpha_{mn} = \sum_j (g^{(j)}_{mn}-g^{(j)}_{my}g^{(j)}_{yn}/g^{(j)}_{yy})\delta_j$, $\quad \beta_{yn} =-\sum_j (g^{(j)}_{yn}/g^{(j)}_{yy})\delta_j$, $\quad \beta_{yy} = \sum_j (1/g^{(j)}_{yy})\delta_j$, $\quad \gamma_{my}=\sum_j (g^{(j)}_{my}/g^{(j)}_{yy})\delta_j$, $\quad \delta_j=d_j/L$ are geometric factors, and $m,n=x,z$.

Expressing $\langle Q_y\rangle$ from the second equation in system (\ref{eq:mean_2}) and substituting it into the rest two equations, the constitutive equations for the flux densities of the effective medium $\langle\vec Q\rangle = \hat g_{eff} \langle\vec P\rangle$ can be derived, where $\hat g_{eff}$ is a tensor
\begin{equation}
\hat g_{eff}=\left( {\begin{matrix} {\tilde \alpha_{xx}} & {\tilde
\gamma_{xy}} & {\tilde \alpha_{xz}} \cr {\tilde \beta_{yx}} &
{\tilde \beta_{yy}} & {\tilde \beta_{yz}} \cr {\tilde \alpha_{zx}} &
{\tilde \gamma_{zy}} & {\tilde \alpha_{zz}} \cr
\end{matrix}
} \right) \label{eq:gem}
\end{equation}
with components $\tilde \alpha_{mn} = \alpha_{mn}-\beta_{yn}\gamma_{my}/\beta_{yy}$, $\tilde \beta_{yn} = -\beta_{yn}/\beta_{yy}$, $\tilde \beta_{yy} = 1/\beta_{yy}$, and $\tilde \gamma_{my} = \gamma_{my}/\beta_{yy}$.

Taking into account nonzero components of relative permittivity and relative permeability tensors of underlying ferrite and semiconductor subsystems, both the effective permittivity tensor $\hat\varepsilon_{eff}$ and the effective permeability tensor $\hat\mu_{eff}$ of the composite finely-stratified ferrite-semiconductor structure can be obtained from (\ref{eq:gem}) via the substitutions $g_{xx} = g^{(f)}_{xx}\delta_f+g^{(s)}_{xx}\delta_s+ (g^{(f)}_{xy}-g^{(s)}_{xy})^2\delta_f\delta_s\Gamma$, $\quad g_{xy} = -g_{yx} = (g^{(f)}_{xy}g^{(s)}_{yy}\delta_f + g^{(s)}_{xy}g^{(f)}_{yy}\delta_s)\Gamma$, $\quad g_{yy} = g^{(f)}_{yy}g^{(s)}_{yy}\Gamma$, $\quad g_{zz}=g^{(f)}_{zz}\delta_f+g^{(s)}_{zz}\delta_s$, $\quad g_{xz} = g_{zx}= g_{yz}= g_{zy} = 0$, and $\Gamma = (g^{(s)}_{yy}\delta_f+g^{(f)}_{yy}\delta_s)^{-1}$. In general there is $g_{xx}\ne g_{zz}$, which means that the considered composite medium is a \textit{biaxial bigyrotropic} crystal.

\section{\label{sec:constpar}Constitutive parameters of ferrite and \\ semiconductor subsystems}

The expressions for tensors components of the underlying constitutive parameters of magnetic $\hat \mu^f\to\hat g^{(f)}$ and semiconductor $\hat \varepsilon^s\to\hat g^{(s)}$ layers can be written in the form
\begin{equation}
\hat g^{(j)}=\left( {\begin{matrix}
   {g_1} & {ig_2} & {0} \cr
   {-ig_2} & {g_1} & {0} \cr
   {0} & {0} & {g_3} \cr
\end{matrix}
} \right). \label{eq:gfs}
\end{equation}

For magnetic layers \cite{Gurevich_book_1963, Collin_book_1992} the components of tensor $\hat g^{(f)}$ are $g_1=1+\chi' + i\chi''$, $\quad g_2=\Omega'+i\Omega''$, $ g_3=1$, and $\quad\chi'=\omega_0\omega_m[\omega^2_0-\omega^2(1-b^2)]D^{-1}$, $\chi''=\omega\omega_m b[\omega^2_0+\omega^2(1+b^2)]D^{-1}$, $\quad\Omega'=\omega\omega_m[\omega^2_0-\omega^2(1+b^2)]D^{-1}$, $\Omega''=2\omega^2\omega_0\omega_m bD^{-1}$, $\quad D=[\omega^2_0-\omega^2(1+b^2)]^2+4\omega^2_0\omega^2 b^2$, where $\omega_0$ is the Larmor frequency and $b$ is a dimensionless damping constant.

For semiconductor layers \cite{Bass_book_1997} the components of tensor $\hat g^{(s)}$ are $g_1=\varepsilon_l\left[ {1-\omega_p^2 (\omega+i\nu)[\omega((\omega+i\nu)^2-\omega_c^2)]^{-1}}\right]$, $\quad g_2=\varepsilon_l\omega_p^2\omega_c[\omega((\omega+i\nu)^2-\omega_c^2)]^{-1}$, $\quad g_3=\varepsilon_l\left[{1-\omega_p^2[\omega(\omega+i\nu)]^{-1}}\right]$, where $\varepsilon_l$ is the part of permittivity attributed to the lattice, $\omega_p$ is the plasma frequency, $\omega_c$ is the cyclotron frequency and $\nu$ is the electron collision frequency in plasma.

Relative permittivity $\varepsilon^f$ of the ferrite layers as well as relative permeability $\mu^s$ of the semiconductor layers are scalar quantities.


\bibliographystyle{elsarticle-num}
\bibliography{Tuz_surface}






\end{document}